\RequirePackage{fix-cm}
\documentclass[smallextended]{svjour3}       
\smartqed  
\usepackage{amssymb,amsmath}
\usepackage{graphicx}
\newcommand{\ket}[1]{\vert {#1} \rangle}  
\newcommand{\C}{\mathbb{C}}

\begin{document}

\title{Towards a Multi Target Quantum Computational Logic
}


\author{Giuseppe Sergioli          
}


\institute{Giuseppe Sergioli \at
              University of Cagliari, Via Is Mirrionis 1, I-09123 Cagliari, Italy\\
              \email{giuseppe.sergioli@gmail.com}           
}

\date{Received: date / Accepted: date}

\maketitle

\begin{abstract}
Unlike the standard Quantum Computational Logic (QCL), where the carrier of information (target) is conventionally assumed to be only the last qubit over a sequence of many qubits, here we propose an extended version of the QCL (we call Multi Target Quantum Computational Logic, briefly MT-QCL) where the number and the position of the target qubits are arbitrary.
\end{abstract}

\section{Introduction}
Both in classical and in quantum computation, a circuit is described in terms of a sequence of gates that transform an arbitrary state (input) into another state (output) \cite{KSV}. In classical computation these transformations are basically irreversible: the Boolean functions $f$ that represent the classical gates are many-to-one, i.e. $f:\{0,1\}^n\mapsto\{0,1\}$, where the dimension of the input state does not correspond to the dimension of the output. Hence, the same output may be obtained in correspondence with different inputs, producing an irreversible process. Anyway, this irreversibility is not longer essencial. As proved by Toffoli \cite{T}, through a very simple expedient any classical irreversible gate can be canonically converted into a respective reversible gate. Given an arbitrary irreversible gate, any input can be divided in two components: the first is the ``genuine" input while the second is an \emph{ancilla} (or \emph{target} bit); on the other hand, also the output can be divided in a first component, which is a copy of the ``genuine" input, and a second component that is obtained by the evolution of the ancilla. Basically, only the second part of the output contains the result of the computation, while the first part only contains redundant information, that represents a kind of \emph{memory} of the input state. Using this expedient, the dimensions of the input and the output states are the same and the computation becomes reversible. The price to pay in order to obtain the desirable reversibility is an increasing of the dimension of the computational space. 

Unlike classical computation, the theory of quantum computation is naturally reversible \cite{SGP}. The Schr\"odinger equation describes the dynamic evolution of quantum systems, showing how the state $\ket{\psi_{t_0}}$ at the initial time $t_0$ evolves into another state $\ket{\psi_{t_1}}$ at the final time $t_1$ by the equation $\ket{\psi_{t_0}}\rightarrow \ket{\psi_{t_1}}=U\ket{\psi_{t_0}}$, where $U$ is a unitary operator that represents a reversible transformation. The Schr\"odinger equation is naturally applied also in quantum information theory \cite{NC}. Indeed, quantum logical gates are unitary operators that describe the time evolution of a quantum input state into an output and to perform a quantum algorithm exactly means to apply a sequence of quantum gates to a quantum input state. Unlike the classical case, because of their unitarity, the dimensions of the input and the output of a quantum gate are the same and the quantum logical gate always represents a reversible transformation. 
Furthermore, another peculiar difference between classical and quantum information theory is given by the basic information quantity, that, in the classical framework, is stored by the classical bit while in quantum computation is given by the quantum bit (\emph{qubits} for short), i.e. unitary vector in the complex Hilbert space 
$\C^2$ \cite{DGS,NC}.
 The qubits turns out to store a very larger amount of information with respect to its classical counterpart and this is the reason that makes, in principle, quantum computation more efficient with respect to the classical one. The input of a quantum circuit is given by a composition of qubits that is mathematically represented by the tensor product operation. Hence, given $k$ qubits $\ket {x_1},\ket {x_2},\cdots,\ket {x_k}\in\C^2$ the input state given by an ensemble of $k$ qubits is given by $\ket {x_1}\otimes\ket {x_2}\otimes\cdots\otimes\ket {x_k}$ (that, for short, we call quantum register - or \emph{quregister} - and we indicate by $\ket {x_1 x_2 \cdots x_k}\in\otimes^k\C^2$). It is important to remark that, given the non-commutativity of the tensor product, the sequence in which any qubit appears in the state is not negligible; in other words, $\ket{x_1 x_2}$ and $\ket{x_2 x_1}$ generally represent two different states.

A quantum circuit is represented by the evolution of the input quregister under the application of some unitary quantum logical gates \cite{H,NC}. Obviously, it is often the case where a $n$-ary quantum gate $U^{(n)}$ is applied to $n$ qubits over a given input configuration, where these $n$ qubits can also be not \emph{adjacent} one another. This very common problem is related to the topic of the architecture in quantum computer design, that plays a crucial role for the realization of advanced technologies in quantum computation \cite{Linke2017}. Even if physical architectures conveniently use particular constraints on the qubits distribution based on the nearest-neighbor couplings \cite{K,Linke2017}, in principle, these constraints have not incidence in the possibility to perform arbitrary computations, because the $Swap$ operations can be suitably used; anyway, the use of the $Swap$ operations is not free of any computational cost. 
On this basis, recent topics related to efficient quantum computing between remote qubits in nearest-neighbor architectures - such as the linear neighbor architectures (LNN) \cite{K} - are active and important areas of research, also devoted to physical implementations \cite{FHH,TKO}.\footnote {As an example, the linear neighbor architectures LNN \cite{K}, offer an appropriate approximate method to approach to physical problems regarding trapped ions \cite{HHRB}, liquid nuclear magnetic resonance \cite{LSB} and the original Kane model \cite{Ka}.}

In addition, from a more theoretical point of view, the architecture in quantum computer design plays also a crucial role in the very general problem regarding the classical simulation of quantum circuits. As focused by Jozsa and Miyake \cite{JM}, the capability of a classical computer to efficiently simulate a quantum circuit is strictly related to the ``distance" between the qubits on which a given quantum gate operates, i.e. the number of the $Swap$ operators necessary to simulate the circuit. 

On this basis, in \cite{S} a simple mathematical representation of an arbitrary quantum circuit is provided. This representation turns out to be very beneficial for pratical usages (for istances related to the implementation of a software package able to efficiently simulate a quantum circuit \cite{Gerdt1,Gerdt2}). But, in addition to the implementative purposes, this representation  could be very useful also to provide a generalization of the quantum computational logic that takes into account a more realistic scenario. This paper is devoted to introduce a new version of the quantum computational logic that is strictly dependent to the architecture of the quantum circuit and that, at the same time, represents a generalization of the standard quantum computational logic \cite{DGG}. 

The paper is organized as follows: in Section 2 we discuss on the different roles that can assume different qubits along a computation, depending on the architecture of the quantum circuit. In Section 3 we briefly describe a formal model to represent an arbitrary quantum circuit by using a synthetic block matrix representation that allows to avoid the reiterate involvement of multiple $Swap$ operators. Section 4 is devoted to introduce a very fundamental background of the main features of the standard Quantum Computational Logic (QCL). Sections 5 and 6 constitute the real core of this paper: in Section 5 the so called Multi Target Quantum Computational Logic (MT-QCL) is introduced in details, while in Section 6 similarities and differences between QCL and MT-QCL are formally showed. The closing section is devoted to introduce some final comments and possible further developments.

\section{Different roles of the qubits in a quantum circuit}
In the standard computational framework, a quantum circuit can be described by three main ingredients: an input state, a logical gate (or a sequence of logical gates) and an output state. In the quantum computational context, the input state is a $n$-dimensional unitary vector $\ket x$ belonging to the complex Hilbert space $\otimes^n\mathbb C^2$, the logical gates are unitary operators and the output state is again a $n$-dimensional unitary vector $\ket y\in\otimes^n\mathbb C^2.$ Of course, the input state is arbitrary and the output is univocally determined by the application of the gate to the input state; indeed, the quantum circuit is identified just as a sequence (i.e. a product) of unitary operators (i.e. one unitary operator obtained as a product of some unitary operators).

The input of a certain quantum circuit can be given by one qubit only or, more frequently, by many qubits. Let us consider an input state $\ket x=\ket{x_1}\otimes\ket{x_2}\dots\ket{x_n}=\ket{x_1\dots x_n}.$ Depending on the gates that are applied during the computation, not all the qubits play the same role. As an example, a very useful class of gates - used for many algorithmic implementations - is given by the following class of \emph{Controlled-U} gates:

$$CU^{(n)}\ket{x_0\cdots x_n}=
\begin{cases}
\ket{x_0\dots x_n} & \mbox{if} \hspace{0.1cm} x_0=0;\\
\ket{x_0}\otimes U^{(n)}\ket{x_1\dots x_n} & \mbox{if} \hspace{0.1cm} x_0=1;
\end{cases}$$

where $U^{(n)}$ is an arbitrary $n$-dimensional unitary operator. 

From a logical perspective, a $CU^{(n)}$ gate can be interpreted as a kind of ``\emph{double implication}" that forces the application of $U^{(n)}$ on $\ket{x_1\dots x_n}$ if and only if $x_0=1$. For this reason, we conventionally say that $\ket 0$ plays the role of the ``\emph{control}" qubit (that keeps unchanged under the application of the $CU^{(n)}$ gate), while $\ket{x_1}, \ket{x_1},\dots,\ket{x_n},$ are called ``\emph{target}" qubits (that, accordingly to the value of $\ket 0$, can change under the application of the $CU^{(n)}$ gate).

It is not hard to show \cite{SF} that, for any $n$-ary unitary operator $U^{(n)}$, the respective $CU^{(n)}$ operator (whose dimension is $n+1$) assumes the following useful matrix representation:

\begin{eqnarray}\label{CU}
CU^{(n)} 
&=& \left[
\begin{array}{c|c}
I^{(n)} &	0 	\\ \hline
0	& 	U^{(n)}
\end{array} 
\right]=
\left[
\begin{array}{c|c}
I^{(n)} &	0 	\\ \hline
0	& 	0
\end{array} 
\right]
+
\left[
\begin{array}{c|c}
0	&	0 	\\ \hline
0	& 	U^{(n)}
\end{array} 
\right]=\\
&=&
P_0 \otimes I^{(n)}+P_1 \otimes U^{(n)},
\end{eqnarray}
where $P_0$ and $P_1$ are the projector operators: $P_0=\ket 0 \langle 0| =\left[ \begin{array}{cc}
					 1 & 0  \\    
					 0 & 0  \\ 
            
						\end{array}\right]$ and $P_1=\ket 1 \langle 1|=\left[ \begin{array}{cc}
					 0 & 0  \\    
					 0 & 1  \\ 
						\end{array}\right].$

Let us remark how a qubit can be neither a control nor a target qubit. 
Let us consider the following picture.

\begin{figure}[htbp]\label{Binary}
\centering
\includegraphics[scale=.18]{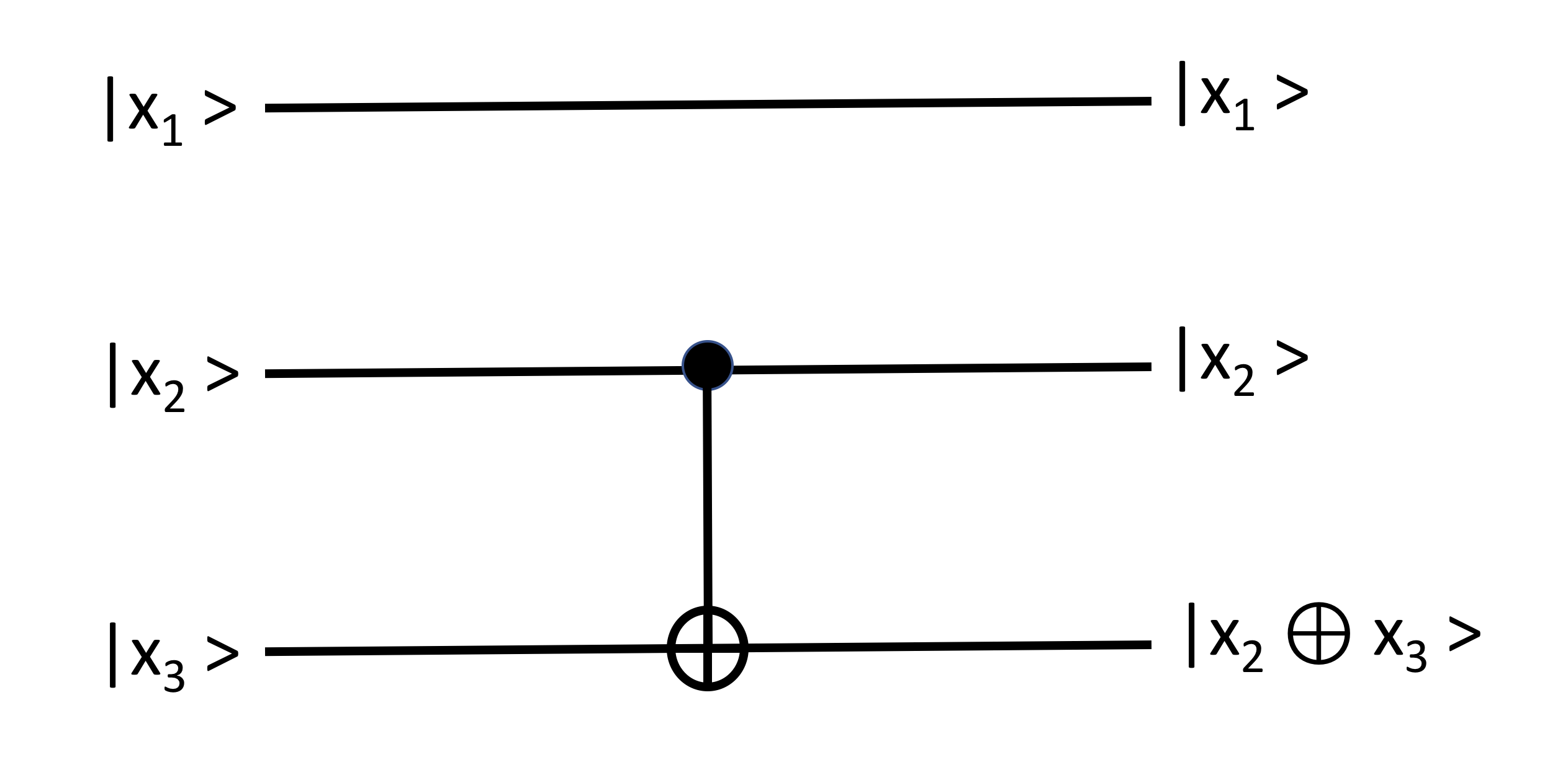}
\caption{Different roles of a qubit in a quantum circuit}
\end{figure}

In this picture the line between the second and the third qubit follows the standard representation of a $CNot$ gate, where the qubit $\ket {x_2}$ plays the role of the control qubit, $\ket {x_2}$ plays the role of the target qubit and the unitary operator $U^{(n)}$ is represented by the standard negation $Not^{(1)}$. Further, the circuit also includes the qubit $\ket{x_1}$ that does play any role in the computation. Formally, the circuit can be represented as the following transformation:
$$(I^{(1)}\otimes CNot)\ket{x_1 x_2  x_3}=\ket{x_1 x_2 (x_2\oplus x_3)},$$ where $I^{(1)}$ is the one dimensional identity matrix and $\oplus$ is the sum modulo $1$. In this case, we can realize the following three different cases: $i)$ $\ket {x_1}$ is a qubit that is left unchanged during the computation and has not incidence in the evolution of the other qubits (it is neither a control nor a target qubit); $ii)$ $\ket {x_2}$ is a control qubit, i.e. it is left unchanged during the computation but it has a relevant incidence in the evolution of the third qubit; $iii)$ $\ket{x_3}$ is a target qubit and, accordingly to the value of $\ket{x_2}$, it can change during the computation.

As a further remark, let us notice that a qubit is generally not designed, in principle, to be a control, a target or neither a control nor a target qubit. Rather, appling a sequence of different quantum gates $U_1,\dots,U_n$ it is simply possible that a given qubit plays a role under the action of $U_i$ but another role under the action of $U_j$. This point will be remarked in the next sections.

\section{Generalized Quantum Circuits}\label{Circuit}
In the previous section particular attention has been devoted to the role of the $Controlled-U$ gates in the architecture of a quantum circuit. In this section we briefly provide the mathematical description of arbitrary quantum circuits, where arbitrary gates are involved. This representation turns out to be useful in the rest of the paper.

The simplest case is represented by a circuit where only a single unary gate is applied. However, this information is not enough to mathematically describe the quantum circuit; indeed, a crucial information is regarding the qubit to which the gate is applied to. Let us consider a unary gate $U^{(1)}$ applied to the $i$-th qubit over a $k$-dimesional circuit (i.e. a circuit such that the input and the output live in the Hilbert space $\otimes^k\mathbb C^2$). In this case, the circuit is simply represented by the following operator:
$$I^{(i-1)}\otimes U^{(1)}\otimes I^{(k-i)}.$$

Similarly, it is possible that $j$ unary operators $U^{(1)}_1,\dots,U^{(1)}_j$ are applied to different $(i_1)^{st},\dots,(i_j)^{th}$ qubits, respectively, over a $k$-dimensional input state. In this case, the circuit is formally  represented as:
\begin{eqnarray}\label{operator}
U=I^{(i_1-1)}\otimes U_1^{(1)}\otimes I^{(i_2-i_1-1)}\otimes U_{2}^{(1)}\otimes\cdots\otimes U_j^{(1)}\otimes I^{(k-i_{j})}.
\end{eqnarray}
Anyway, a more complex quantum circuit is generally given by sequences of gates in such a way that any qubit can change under the action of several quantum gates. As an example, it is possible to the consider the operator $\tilde{U}$ that corresponds to the application of $\tilde{j}$ unitary operators $\tilde{U}^{(1)},\dots,\tilde{U}^{(1)}_{\tilde j}$ applied to different $(\tilde i_1)^{st},\dots,(\tilde i_{\tilde j})^{th}$ qubits. Given a $k$-dimensional input state $\ket{\psi_{in}}$, it is possible to consider to apply $\tilde U$ not to the input, but to the $k$-dimesnional state that arises from the application of $U$ to the input state $\ket{\psi}$. In this case the circuit is described by the product $\tilde{U}U$ and, trivially, the output is $\ket{\psi_{out}}=\tilde{U}U\ket{\psi}.$

We have considered only the special case of a circuit given by unitary operators only. By a suitable arrangment of the dimension of the identity operators, it is easy to generalize the argument above to the case of circuits given by compositions of gates of different and arbitrary arity. 

As an example, let us consider a $3$-dimensional input state $\ket{\psi}=\ket{x_1x_2x_3}$ where we first apply a $CNot$ between the first and the second qubit and an Hadamard gate $H$ to the third qubit. After, we apply a $CNot$ between the second and the third qubit. In this case, the circuit is given by the following composition:
$$(CNot\otimes H)(I^{(1)}\otimes CNot).$$ This example is also useful to realize how the role of a qubit of the input state is not designed, in principle, to be a control or a target bit; indeed, the qubit $\ket{x_2}$ plays the role of target during the first ``step" of the computation and the role of control during the second ``step".

All the arguments above contain the implicit assumption that a $n$-ary gate is applied to $n$ adjacent qubits. In a more general scenario, a $n$-ary gate $U^{(n)}$ could also be applied to $n$ qubits arbitrary placed over a $k$-dimesional Hilbert space (with $k\geq n$). In this case, the very standard quantum computational procedure requires a full involvment of the $Swap$ operator, by the following strategy.
Let us consider a $k$-dimensional input state $\ket x^{in}=\ket{x_1 \dots x_m \dots x_{m+n} \cdots x_k}$. For the sake of the simplicity, let us consider a binary gate $U^{(2)}$ that is applied to the two non adjacent qubits $\ket{x_m}$ and $\ket{x_{m+n}}$, leaving the other qubits of the input state $\ket x^{in}$ unchanged. 

First, we apply $(n-1)$ $Swap$ gates in order to ``move" the $\ket{x_m}$ in the $(m+n-1)^{th}$ position. Formally: $$\ket{x}^a=Swap_{[k;m,m+n-1]}\ket{x}^{in},$$ 
where $Swap_{[k;m,m+n-1]}$ represents the composition of all the $(n-1)$ $Swap$ gates we need to ``move" the $\ket{x_m}$ in the $(m+n-1)^{th}$ position.
At this stage it is possible to apply the binary operator $U^{(2)}$ to the two qubits $\ket{x_m}$ and $\ket{x_{m+n}}$ (that are now adjacent). Formally (by referring to Eq.\ref{operator}): 
$$\ket {x}^b=(I^{(m+n-2)}\otimes U^{2}\otimes I^{(k-m-n)})\ket {x}^a.$$ 

Finally, we need to apply the inverse\footnote{It is easy to see that $Swap_{[k;m,m+n-1]}^{-1}=Swap_{[k;m,m+n-1]}$ for any value of $k,m$ and $n$.} of $Swap_{[k;m,m+n-1]}$ in order to retreive the initial configuration:
$\ket{x}^{out}=Swap_{[k;m,m+n-1]}\ket{x}^b.$

We can formally summarinzing this procedure by writing:

$$\ket{x}^{out}=U^{(2)}_{[k;m,m+n]}\ket x^{in},$$

where $U^{(2)}_{[k;m,m+n]}$ indicates the $k$-dimensional operator that acts as a binary operator $U^{(2)}$ on the $m^{th}$ and the $(m+n)^{th}$ qubits of $\ket x^{in}$ an leaves all the other qubits of $\ket x^{in}$ unchanged; the extended representation of $U^{(2)}_{[k;m,m+n]}$ is now simply obtainable by composition of $Swap$, $I$ and $U^{(2)}$ gates, suggested by the previous description.
Without any lost of generality, all the agument above can be easily generalized for the case of $n$-ary operators applied to $n$ qubits arbitrarily allocated over a $k$ dimensional input state. Of course, the ``distance" between the qubits where the gates are applied, the ariety of these gates and the number of these gates are elements that could provide a relevant increasing of the number of the $Swap$ gates necessary to make the required computation. This problem is fully investigated in the general context of architecture in quantum computer design. Indeed, recent topics related to efficient quantum computing among remote qubits in nearest-neighbor architectures are currently under investigation \cite{K,TKO}.
From a purely mathematical viewpoint, it is easy to realize how the using of multiple $Swap$ gates turns out to be particularly unconfortable. For this reason we provide the following block matrix representation \cite{S} that are useful in order to formally manage quantum circuits with different gates applied to non adjacent qubits.

\begin{theorem}\label{Swap}\cite{S}

Let us consider the projectors operators
 $P_0$ and $P_1$ and the Ladder operators \cite{F}
$L_0=\ket 0 \langle 1| =\left[ \begin{array}{cc}
					 0 & 1  \\    
					 0 & 0  \\ 
            
						\end{array}\right]$ and  $L_1=\ket 1  \langle 0| =\left[ \begin{array}{cc}
					 0 & 0  \\    
					 1 & 0  \\ 
            
						\end{array}\right]$ and let $P_0^{(n)}=I^{(n-1)}\otimes P_0$ (with $I^{(n-1)}$ $(n-1)$-dimensional identity matrix); similarly for $P_1^{(n)}, L_0^{(n)}$ and $L_1^{(n)}.$ 

The block-matrix representation of the operator  $Swap_{[k;m,n+n]}$ (that swaps the $m^{th}$ qubit with the $(m+n)^{th}$ qubit over a $k$-dimensional input state) is given by:
$$Swap_{[k;m,m+n]}=I^{(m-1)}\otimes Swap_{[n+1;1,n+1]}\otimes I^{(k-m-n)}$$ where 
$$Swap_{[n;1,n]}=\left[ \begin{array}{c|c}
					P_0^{(n-1)} & L_1^{(n-1)} \\ \hline     
					 L_0^{(n-1)} & P_1^{(n-1)} \\ 
						\end{array}\right].$$
\end{theorem}

This Theorem allows us to provide also a confortable block matrix representation of an arbitrary binary gate $U^{(2)}$ (applied to two non adjacent qubits) without incurring in the annoying involvement of multiple $Swap$ gates.

\begin{theorem}\label{binary}\cite{S}

Let $U^{(2)}$ a binary unitary operator given by the following block-matrix representation $
U^{(2)} =  \left[ \begin{array}{c|c}
					U_{11} & U_{12} \\ \hline          
					  U_{21}&  U_{22} \\ 
						\end{array}\right],
$ where $U_{ij}$ are $2-$ dimensional square matrices given by $U_{11} =  \left[ \begin{array}{cc}
					u_{11} & u_{12} \\           
					  u_{21}&  u_{22} \\ 
						\end{array}\right],$ $ U_{12} =  \left[ \begin{array}{cc}
					u_{13} & u_{14} \\           
					  u_{23}&  u_{24} \\ 
						\end{array}\right]$ , $ U_{21} =  \left[ \begin{array}{cc}
					u_{31} & u_{32} \\           
					  u_{41}&  u_{42} \\ 
						\end{array}\right]$ and $ U_{22} =  \left[ \begin{array}{cc}
					u_{23} & u_{24} \\           
					  u_{43}&  u_{44} \\ 
						\end{array}\right].$

The block-matrix representation of $U^{(2)}_{[k;m,m+n]}$ is given by: $$U^{(2)}_{[k;m,m+n]}=I^{(m-1)}\otimes \left[ \begin{array}{c|c}
					U_{11}^{(n)} & U_{12}^{(n)} \\  \hline         
					  U_{21}^{(n)}&  U_{22}^{(n)} \\ 
					\end{array}\right]\otimes I^{(k-m-n)},$$
where $U_{ij}^{(n)}=I^{(n-1)}\otimes U_{ij}.$
\end{theorem}

It is not hard to convince that, by following very similar arguments, it is possible to obtain a similar result for an arbitrary $n$-ary gate applied to $n$ non adjacent qubits \cite{S}.

\section{The standard approach to the Quantum Computational Logic}\label{QCL}

The theory of Quantum Computation has naturally inspired new forms of quantum logic, the so called \emph{Quantum Computational Logic} (QCL) \cite{DGG,SGP}. From a semantic point of view, any formula of the language in the QCL denotes a piece of quantum information, i.e. a density operator living in a complex Hilbert space whose dimension depends on the linguistic complexity of the formula. Similarly, the logical connectives are interpreted as special examples of quantum gates. Accordingly, any formula of a quantum computational language can be regarded as a logical description of a quantum circuit. 
The initial concept at the very background of the QCL is the assignment of the truth value of a quantum state that represents a formula of the language. Conventionally, the QCL assumes        to assign the truth value ``false" to the information stored by the qubit $\ket 0$ and the truth value ``true" to the qubit $\ket 1.$ Unlike the classical logic, QCL turns out to be a \emph{probabilistic} logic, where the qubit $\ket\psi=c_0\ket 0 +c_1\ket 1$ logically represents a ``probabilistic superposition" of the two classical truth values, where the \emph{falsity} has probability $|c_0|^2$ and the \emph{truth} has the probability $|c_1|^2$. As in the qubit case, in the standard approach of QCL it is also defined a probability function $\texttt p$ that assings a probability value $\texttt p(\rho)$ to any density operator $\rho$ living in the space of the $n$-arbitrary dimensional density operators (we denote this space by $\mathcal D(\otimes^n\mathbb C^2)$). Intuitively, $\texttt p(\rho)$ is the probability that the quantum information stored by $\rho$ corresponds to a \emph{true} information. 

In order to define the function $\texttt p$, we first need to identify in any space $\otimes^n\mathbb C^2$ the two operators $P_0^{(n)}$ and $P_1^{(n)}$ as the two special projectors that represent the \emph{falsity} and the \emph{truth} properties, respectively. Before this, a step is very crucial. In order to extend the definition of \emph{true} and \emph{false} from the space $\mathbb C^2$ of the qubits to the space $\otimes^n\mathbb C^2$ of the tensor product on $n$ qubits (i.e. on an arbitrary \emph{quregister}), the standard approach of the QCL accords with the following convention: a quregister $\ket x=\ket{x_1\dots x_n}$ is said to be it is said to be \emph{false} if and only if $x_n=0$; conversely, it is said to be \emph{true} if and only if $x_n=1$. Hence, the truth value of a quregister only depends on its last component. On this basis, it is natural to define the property  \emph{falsity} (or \emph{truth}) on the space $\otimes^n\mathbb C^2$ as the projector $P_0^{(n)}$ (or $P_1^{(n)}$) onto the span of the set of all \emph{false} (or \emph{true}) registers. Now, accordingly with the Born rule, the probability that the state $\rho$ is \emph{true} is defined as: 
\begin{eqnarray}\texttt p(\rho)=Tr(P_1^{(n)}\rho).
\end{eqnarray}
In QCL the evolution of a quregister is dictated by the application of a unitary operator (that represents a reversible transformation) while the evolution of a density operator is dictated by the application of a quantum operation (that represents a transformation that, in general, is not reversible). Of course, for any quantum gate $U$ there exists the correspondent quantum operation $^\mathcal DU$ that replaces the behaviour of the quantum gate in the context of the density operators (in particular, $^\mathcal DU(\rho)=U\rho U^{\dagger}$), but the other way generally does not hold. 

In the language of the QCL it is usual to distinguish between \emph{semiclassical} gates (called \emph{semiclassical} because, when they are applied to the elements of the computational basis $\mathbb B=\{\ket 0, \ket 1\}$, they replace the behaviour of their corresponding classical logical gates) and \emph{genuinely} quantum gates (called \emph{genuinely} quantum because their application to the elements of the computational basis has not any classical counterpart). The semiclassical gates usually involved in the QCL are: the Identity $I$, the Negation $Not$, the control-negation (or Xor) $CNot$ and the Toffoli gate $T$, while the genuinely quantum gates are: the Hadamard gate $\sqrt I$ (also named square root of the identity) and the square root of the negation $\sqrt{Not}.$
In particular, the $T$ and the $Not$ gates allows to provide a \emph{probabilistic} replacement of the classical logic in virtue of the following properties:
\begin{itemize}
\item $\texttt p(^\mathcal D{Not}(\rho))=1-\texttt p(\rho),\,  \text{for any} \rho\in\otimes^n\mathbb C^2;$
\item $\texttt p(AND(\rho,\sigma))=\texttt p(^\mathcal D{T}(\rho\otimes\sigma\otimes P_0))=\texttt p(\rho) \texttt p(\sigma), \, \text{for any} \rho,\sigma \in \otimes^n\mathbb C^2.$
\end{itemize}

Let us notice how the conjunction is obtained by the expedient to use the ternary Toffoli gate equipped by the projector $P_0$ that plays the role of an \emph{ancilla}. 

Basing on this approach and inspired by the intrinsic properties of the quantum systems, the semantic of the QCL   turns out to be strongly non-compositional and context dependent \cite{DGLS}. This approach, that may appear \emph{prima facie} a little strage,   leads to the benefit to reflect pretty well plenty of informal arguments that are currently used in our rational activity \cite{DGLNS}. A  detailed description of the QCL and its algebraic properties are summarized in \cite{DGG,DGLS,DGS,LS}.

\section{A new definition of probability in QCL}
Following the brief description of a quantum circuit given in Section \ref{Circuit}, we can easy realize that the actual carrier of information is given by the target bit, while the control bit remains unchanged under the application of a given quantum gate; furthermore, the set of the qubits that play the role of target can change during the computation, depending on the gates that, time by time, are applied. 
On the other hand, despite its remarkable expressive power, the very preliminary notions of the QCL introduced in Section \ref{QCL} seem to do not take into suitable account this fact, assuming that the ``useful" information is always stored by the last qubit only. Indeed, the assignment of the truth value of a given composition of qubits (a quregister), accordingly with what we have previously introduced, only depends on the last component. For this reason, all the gates that are involved in the language of the QCL are only one-target gates. This restriction is basically unnecessary and it could also seem to be a little far from the architecture of a real quantum circuit. For this reason, this section is devoted to introduce an extension of the QCL (that we will call \emph {Multi Target} QCL, briefly MT-QCL) and to show some preliminary result. The immediate benefit of the MT-QCL with respect to the QCL is given by the fact that the MT-QCL can involve in the language also multi target gates (such as the $Swap$ gate, the square root of $Swap$ gate $\sqrt{Swap}$ and the Fredkin gate $F$) without any lost of generality. Further, in this framework the standard QCL can be seen as a particular (one-target) case of the MT-QCL.

Similarly to the case of QCL, in order to introduct the MT-QCL the essencial step is given by the definition of probability. First, let us consider a very simple circuit given by a $n$-dimensional input state $\ket x= \ket{x_1\dots x_n}$ and only one operator $U^{(n)}$ acting on the space $\otimes^n\mathbb C^2$ as $U^{(n)}\ket{x}=\ket{y_1 \dots y_n}$. Let us consider the two following sets of indexes strictly dependent on the operator $U^{(n)}$: $$C_{U^{(n)}}=\{i:\ket{x_i}=\ket{y_i}\} \,\,\, \text{and} \,\,\, T_{U^{(n)}}=\{j:\ket{x_j}\neq\ket{y_i}\}.$$

Intuitively, $C_{U^{(n)}}$ selects the position of the qubits of the input state that are not affected by $U^{(n)}$; conversely for $T_U^{(n)}.$ Conveniently, let us call any $i$ belonging to $C_{U^{(n)}}$ a \emph{control position} and any $j$ belonging to $T_{U^{(n)}}$ a \emph{target  position}.\footnote{Let us give a slight abuse of the terms \emph{target} and \emph{control} remarking that, for example, if we refer to the picture of the Section \ref{Circuit}, the qubit $\ket{x_1}$ is not a control qubit but, accordingly with this notation, it assumes a \emph{control position} over the quantum circuit where it is located.}
On this basis, we define a probability $\texttt P$ related not to the state only but also to the operator $U^{(n)}$, i.e. we associate a probability to the couple $[U^{(n)},\ket x]$ by the following definition: 

\begin{definition}\label{Prob}[Probability in MT-QCL]  

$$\texttt P[U^{(n)},\ket x]=Tr(\mathcal P_1\otimes\cdots\otimes\mathcal P_n)^{\mathcal D}U^{(n)}\rho_{\ket x}$$
where $U^{(n)}$ is a $n$-dimensional unitary operator, $\rho_{\ket x}=\ket x\langle x|$ and

$$\mathcal P_i=\begin{cases}
I \,\,\,\, \text{if} \,\,  i\in C_{(U^{(n)})}; \\
P_1 \,\,\,\, \text{if} \,\,  i\in T_{(U^{(n)})}.
\end{cases}$$


\end{definition}

Let us remark that the quantity $\mathcal P_1\otimes\dots\otimes\mathcal P_n$ is univocally determined by the operator $U^{(n)}$; in any case, the quantity $\mathcal P_1\otimes\dots\otimes\mathcal P_n$ is a projector operator for any $U^{(n)}$; hence, $\texttt P[U^{(n)},\ket x]$ is a well defined probability.
Further, let us notice that, even if the definition is given in the special case where the input state is pure, the definition can be naturally generalized, without any lost of generality, to the case where the input state $\rho$ is also a mixed state (formally assuming the form $\texttt P[^{\mathcal D}U^{(n)},\rho]$). 

Let us also notice that the identity operator plays a very special rule in this framework, in virtue of the following Theorem.
\begin{theorem}\ \\

$\texttt P[^{\mathcal D}I^{(n)},\rho]=1$ for any $\rho\in\mathcal D(\mathbb C^2).$
\begin{proof}\ \\
Trivially follows by Def.(\ref{Prob}). 
\end{proof}
\end{theorem}
As discussed in Section \ref{Circuit}, when more quantum gates are involved along a computation, it is possible that the same qubit plays, as an example, the role of target during one ``step" of the computation and the role of control during another ``step". Resonably, we accord to fix that if a qubit $\ket{x_i}$ assumes a target position at list one time during the computation, then the position $i$ will be considered as a target position in the evaluation of the probability given by Def. \ref{Prob}.

On this basis, it is possible to naturally generalize the Def. \ref{Prob} to the case where $U^{(n)}$ is not only one gate quantum circuit, but it is an arbitrary circuit given by an arbitrary composition of quantum gates. 
\begin{example}\ \\

Let us consider a quantum circuit given by the following composition of gates
$$U^{(3)}=(CNot\otimes H)(I^{(1)}\otimes CNot).$$ and let $\rho^{in}=\rho\otimes\sigma\otimes\tau$, where $\rho=\frac{1}{2}\left[ \begin{array}{cc}
					 1+r_3 & r_1-ir_2  \\    
					 r_1+ir_2 & 1-r_3  \\ 
						\end{array}\right],$ 

$\sigma=\frac{1}{2}\left[ \begin{array}{cc}
					 1+s_3 & s_1-is_2  \\    
					 s_1+is_2 & 1-s_3  \\ 
						\end{array}\right]$ and $\tau=\frac{1}{2}\left[ \begin{array}{cc}
					 1+t_3 & t_1-it_2  \\    
					 t_1+it_2 & 1-t_3  \\ 
						\end{array}\right].$

We observe that only the first qubit $\rho$ never assumes a \emph{target position}. Hence, by Def. \ref{Prob}, we have that:\footnote{Notice that the probability $\texttt p$ we refer to is the one introduced in the context of the standard QCL (see Eq. 4.1). On this basis, we are using this example to express the MT-QCL probability $\texttt P$ in terms of the QCL probability $\texttt p$.}
\end{example}
$$\texttt P[U^{(3)}, \rho^{in}]=Tr(I^{(1)}\otimes P_1\otimes P_1)^\mathcal DU^{(3)}\rho^{(in)}=$$
$$=\frac{1}{4}[(1-2\texttt p_{\rho})(1-2\texttt p_{\sigma})-1](t_1-1).$$

Let us immediately notice that the main difference between the definition of probability given in the standard QCL and the one given in the MT-QCL is that the first is related to the quantum state, while the second is related to the quantum circuit. Indeed, given a certain output state $\ket x^{out}$, the QCL assignes a probability to $\ket x^{out}$ indipendently to the computation from which the state comes from; conversely, the probability assigned by the MT-QCL is strictly dependent on the \emph{history} of $\ket x^{out}.$ This turns out to be very noticeable by the next example.
\begin{example}\ \\

Let us consider the input state $\rho\in\mathcal D(\otimes^2\mathbb C^2)$ and let also consider the two circuits dictated by the following compositions of quantum gates: $U_1=(\sqrt I\otimes I)(\sqrt I\otimes I)$ and $U_2=I^{(2)}.$ Obviously, $U_1=U_2$ and $^{\mathcal D}U_1(\rho)=^{\mathcal D}U_2(\rho)=\rho$. But, $$\texttt P[^{\mathcal D}U_1,\rho]=Tr(P_1\otimes I)\rho\neq Tr(I\otimes I)\rho=\texttt P[^{\mathcal D}U_2,\rho].$$
Indeed, even if $U_1=U_2$, the first position of $U_1$ is a target position, while the first position of $U_2$ is a control position.
This difference is captured by the  MT-QCL in virtue of the new definition of probability.
\end{example}

\begin{figure}[ht]
\begin{center}
\includegraphics[scale=0.14]{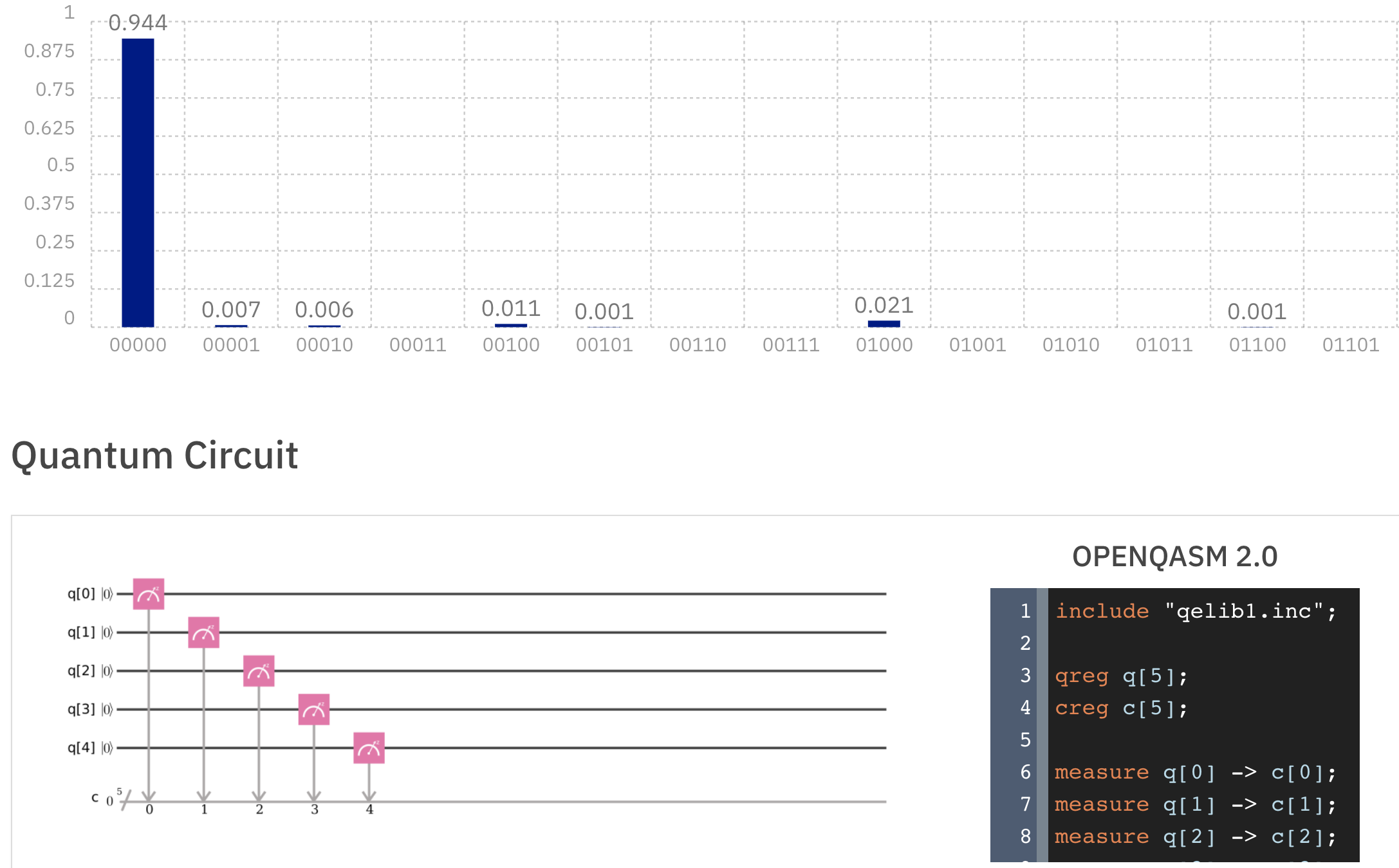}
\includegraphics[scale=0.15]{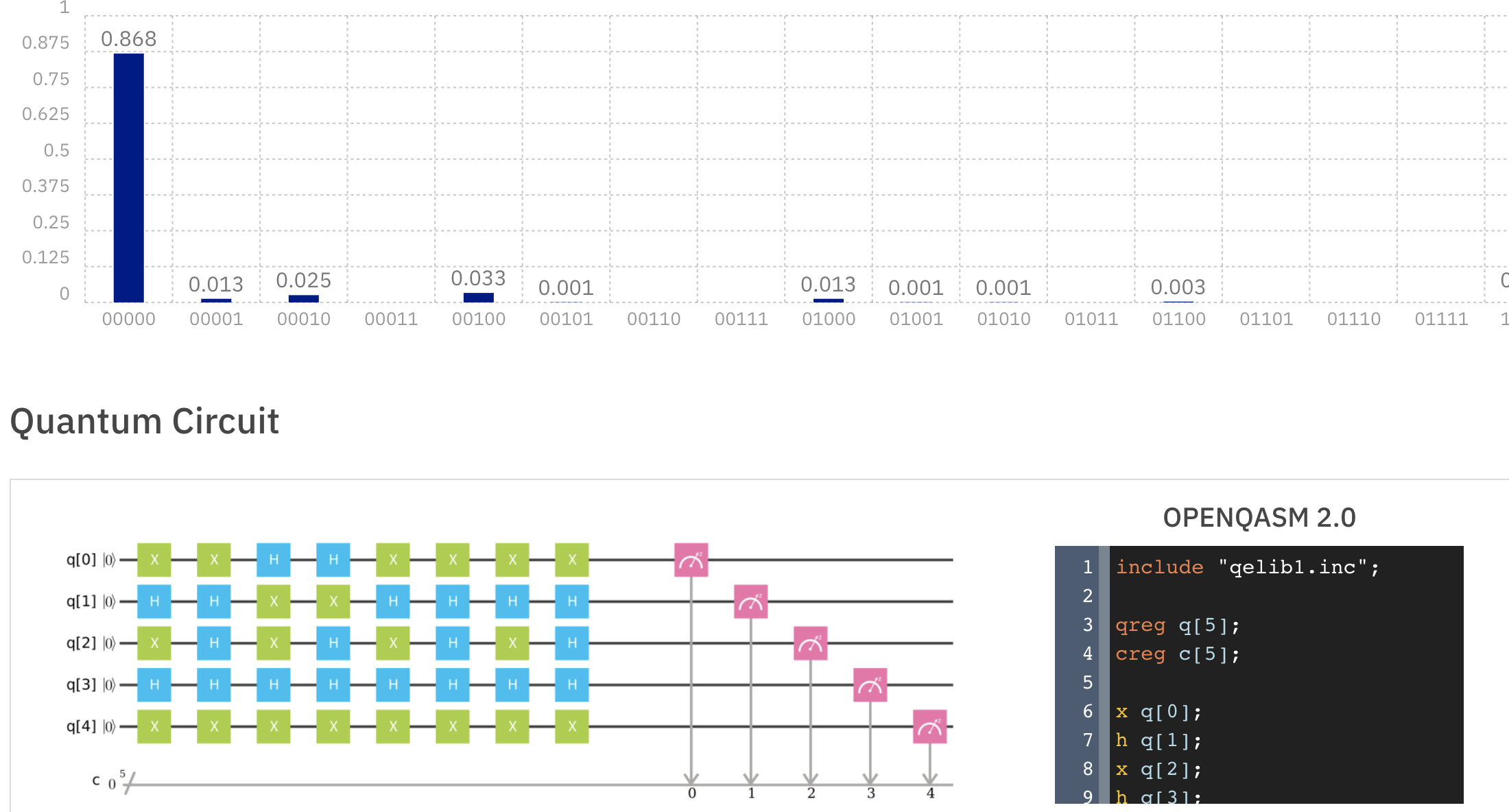}
\caption{\label{diff} Two equivalent circuits with different probability distributions}
\end{center}
\end{figure}

The fact that two (let's say) ``equivalent" but not ``identical" circuits (such as $U_1$ and $U_2$ of the previous example) provide different probabilistic results in the framework of MT-QCL is not so surprising. Indeed, even in the real quantum computation, two equivalent (but not identical) circuits applied to the same input state can provide as output a different distribution. In Fig.\ref{diff} we show the result of the performing of two equivalent (but not identical) identity circuits applied to the same input state $\ket{\psi}=\ket{00000}.$ The probabilistic results is obtained by running the output on a large number of computations on the real IBM quantum computer \cite{De}. As we can see, even if the expected result for both case is, obviously, $\ket{\psi}$, the probabilistic distributions in the two cases are different. This difference arises from the fact that, even if from an operational point of view the two circuits are exactly equivalent, the computations that are executed by the quantum computer are essentially different.  

\section{Some result}
Now let us generalize the notation that we have introduced in Section \ref{Circuit}, and let us indicate by $U{(\alpha)}_{[k;m,m+n_1,\dots,m+n_{\alpha-1}]}$ an $\alpha$-ariety gate that is applied to the $m^{th}, (m+n_1)^{th},\dots,(m+n_{\alpha-1})^{th}$ qubits, respectively, and let calculate the value that assumes the new definition of probability associated to some particular gate.

\begin{theorem}\label{T1}\ \\

Let $\rho_1,\rho_2,\dots,\rho_k$ be density operator leaving in the two dimensional Hilbert space $\mathcal D(\mathbb C^2)$ such that $\rho_i=\frac{1}{2}\left[ \begin{array}{cc}
					 1+r_{i3} & r_{i1}-ir_{i2}  \\    
					 r_{i1}+ir_{i2} & 1-r_{i3}  \\ 
						\end{array}\right]$. We have that:
\begin{enumerate}
\item $\texttt P[^{\mathcal D}Not^{(1)}_{[k;i]},(\rho_1\otimes\cdots\otimes\rho_i\otimes\rho_k)]=\frac{1+r_{i3}}{2};$
\item $\texttt P[^{\mathcal D}\sqrt I^{(1)}_{[k;i]},(\rho_1\otimes\cdots\otimes\rho_i\otimes\rho_k)]=\frac{1- r_{i1}}{2};$
\item $\texttt P[^{\mathcal D}\sqrt {Not}^{(1)}_{[k;i]},(\rho_1\otimes\cdots\otimes\rho_i\otimes\rho_k)]=\frac{1- r_{i2}}{2};$
\item $\texttt P[^{\mathcal D}CNot_{[k;m,m+n]},(\rho_1\otimes\dots\otimes\rho_k)]=\frac{1-r_{m3}s_{(m+n)3}}{2};$
\item $\texttt P[^{\mathcal D}T_{[k;m,m+n,m+n+p]}),(\rho_1\otimes\dots\otimes\dots\rho_k)]=\frac{1}{4}(2+(r_{m3}(r_{(m+n)3}-1)-r_{(m+n)3}-1)\cdot r_{(m+n+p)3}).$
\end{enumerate}

\begin{proof}
 \begin{enumerate}
\item \begin{eqnarray*}
& & \texttt  P[^{\mathcal D}Not_{[k;i]}^{(1)},(\rho_1\otimes\dots\otimes\rho_i\otimes\dots\otimes\rho_k)]=\\
&=& Tr[(I^{(i-1)}\otimes P_1\otimes I^{(k-i)})\cdot\,^{\mathcal D}(I^{(i-1)}\otimes Not\otimes I^{(k-i)})\cdot (\rho_1\otimes\dots\otimes\rho_i\otimes\cdots\rho_k)]=\\
&=&Tr[(I^{(i-1)}\cdot\,^{\mathcal D}I^{(i-1)}(\rho_1\otimes\cdots\otimes\rho_{i-1}))\otimes(P_1\cdot\,^{\mathcal D}Not\,\rho_i)\otimes \\
&\otimes&(I^{(k-i)}\cdot\,^{\mathcal D}I^{(k-i)}(\rho_{i+1}\otimes\cdots\otimes\rho_{k}))]=\frac{1+r_{i3}}{2};
\end{eqnarray*}
\item follows in a similar way; \item follows in a similar way;
\item \begin{eqnarray*}
& & \texttt P[^{\mathcal D} CNot_{[k;m,m+n]},(\rho_1\otimes\dots\otimes\rho_m\otimes\dots\otimes\rho_{m+m}\otimes\dots\otimes\rho_k)]=\\
&=&Tr[(I^{(m+n-1)}\otimes P_1\otimes I^{k-m-n})\cdot\,^{\mathcal D}Swap_{[k;m,m+n-1]}\,(^{\mathcal D}(I^{(m+n-2)}\otimes CNot\otimes I^{(k-m-n)})\cdot \\
&\cdot& (^{\mathcal D}Swap_{[k;m,m+n-1]}(\rho_1\otimes\dots\otimes\rho_k)))]=\\
& & \text{(by Theorem \ref{binary})}\, =Tr(P_1(CNot(\rho_m\otimes\rho_{m+n})))=\frac{1-r_{m3}s_{(m+n)3}}{2};
\end{eqnarray*}
\item follows in a similar way.
\end{enumerate}
\end{proof}
\end{theorem}

Let us notice that the previous results are in accord with the standard QCL \cite{DGS} in the cases (1), (2) and (3). Also for the case (4), after considering to ``move" the qubit $\rho_m$ in the position $(m+n-1)$  by a suitable availment of $Swap$ (in accord with Theorem \ref{Swap}) in order to have $\rho_m$ and $\rho_{m+n}$ in two adjacent positions, we are prefectly recovering the probability values obtained by the standard QCL. It is easy to think in a very similar way regarding the item (5).

As we have discussed in Section \ref{Circuit}, the action of the $Swap$ gate allows to generalize the action of $n$-ary gates to the case where the qubits to which the gates are applied are not adjacent one another. Anyway, it has not any influence in the calculation of the probability; as an example: $\texttt P[^{\mathcal D} CNot_{[k;m,m+n]}),(\rho_1\otimes\cdots\otimes\rho_m\otimes\cdots\otimes\rho_{m+n}\otimes\cdots\otimes\rho_k)]$ it is simply equal to $\texttt P[^{\mathcal D} CNot,(\rho_m\otimes\rho_{m+n})]$: hence, without lost of generality, we can confine ourselves in calculating the probability in the special case in which the qubits where the gate acts on are adjacent.

\begin{theorem}\ \\

Let $\rho_1,...,\rho_k$ as defined in Theorem \ref{T1}.

\begin{enumerate}
\item $\texttt P\,[^{\mathcal D}Swap,(\rho_1\otimes\rho_2)]= \mathcal P(\rho_1)\mathcal P(\rho_2);$
\item $\texttt P\,[^{\mathcal D}\sqrt{Swap},(\rho_1\otimes\rho_2)]= \mathcal P(\rho_1)\mathcal P(\rho_2);$
\item $\texttt P\,[^{\mathcal D}F,(\rho_1\otimes\rho_2\otimes\rho_3)= \mathcal P(\rho_2)\mathcal P(\rho_3)];$
\end{enumerate}

\begin{proof}\ \\

Trivially follow from Definition \ref{Prob} and by straightforward calculations.
\end{proof}
\end{theorem}
Let us notice that, by using the definition of probability provided in the standard QCL, the results of previous theorem should assume remarkable different values.

By the way, the most remarkable results should regard the case where the input state is not a product state. On this basis, many results have been obtained regarding one-target gates \cite{DGLS,SF}. Following similar arguments and according with the new definition of probability, we show similar results also for the $Swap$, $\sqrt{Swap}$ and $F$ gates.

\begin{theorem}\label{binary2}\ \\

Let $U$ a binary operator $U=\left[ \begin{array}{c|c}
					U_{11} & U_{12} \\ \hline     
					 U_{21} & U_{22} \\ 
						\end{array}\right]$ (as represented in Theorem (\ref{binary})) and let us consider that $U$ is not a control-target operator. Let $\rho\in\mathcal D(\otimes^k\mathbb C^2).$ Then:
$$\texttt P[^{\mathcal D}U_{[k;m,m+n]},\rho]=Tr[(\Lambda_U^{(m+n)}\otimes I^{(k-m-n)})\rho],$$ where $\Lambda_U^{(n+1)}=\left[ \begin{array}{c|c}
					I^{(n-1)}\otimes(U_{21}^{\dagger}P_1U_{21}) & I^{(n-1)}\otimes(U_{21}^{\dagger}P_1U_{22})  \\ \hline     
					 I^{(n-1)}\otimes(U_{22}^{\dagger}P_1U_{21}) & I^{(n-1)}\otimes(U_{22}^{\dagger}P_1U_{22})  \\ 
						\end{array}\right].$
\begin{proof}\ \\

By Def.(\ref{Prob}), we have:
\begin{eqnarray*}
& & \texttt P[U_{[k;m,m+n]},\rho]=\\
&=& Tr[(P_1^{(m)}\otimes P_1^{(n)}\otimes I^{(k-m-n)})U_{[k;m,m+n]}\,\,\rho\,\, U_{[k;m,m+n]}^{\dagger}] =\\
&=& Tr[(U_{[k;m,m+n]}^{\dagger}(P_1^{(m)}\otimes P_1^{(n)}\otimes I^{(k-m-n)})U_{[k;m,m+n]})\,\rho]=\\
 & & \text{(by Theorem (\ref{binary}))}
= Tr[(I^{(m-1)}\otimes  \left[ \begin{array}{c|c}
					U_{11}^{(n)\dagger} & U_{21}^{(n)\dagger} \\ \hline     
					 U_{12}^{(n)\dagger} & U_{22}^{(n)\dagger} \\ 
						\end{array}\right] \otimes I^{(k-m-n)})\cdot \\ 
&\cdot& (I^{(m-1)}\otimes (P_1\otimes P_1^{(n)})  \otimes I^{(k-m-n)}) \cdot \\
&\cdot& (I^{(m-1)}\otimes \left[ \begin{array}{c|c}
					U_{11}^{(n)} & U_{12}^{(n)} \\ \hline     
					 U_{21}^{(n)} & U_{22}^{(n)} \\ 
						\end{array}\right]  \otimes I^{(k-m-n)}) \rho]=\\
&=& Tr[(I^{(m-1)}\otimes(\left[ \begin{array}{c|c}
					U_{11}^{(n)\dagger} & U_{21}^{(n)\dagger} \\ \hline     
					 U_{12}^{(n)\dagger} & U_{22}^{(n)\dagger} \\ 
						\end{array}\right]\cdot \left[ \begin{array}{c|c}
					{\bf 0}^{(n)} & {\bf 0}^{(n)} \\ \hline     
					 {\bf 0}^{(n)} & P_1^{(n)} \\ 
						\end{array}\right]\cdot \left[ \begin{array}{c|c}
					U_{11}^{(n)} & U_{12}^{(n)} \\ \hline     
					 U_{21}^{(n)} & U_{22}^{(n)} \\ 
\end{array}\right])\otimes I^{(k-m-n)})\,\rho],
\end{eqnarray*}
where ${\bf 0}^{(n)}$ is the $n$-dimensional null matrix. Let us notice that
\begin{eqnarray*}
&  \left[ \begin{array}{c|c}
					U_{11}^{(n)\dagger} & U_{21}^{(n)\dagger} \\ \hline     
					 U_{12}^{(n)\dagger} & U_{22}^{(n)\dagger} \\ 
						\end{array}\right]\cdot \left[ \begin{array}{c|c}
					{\bf 0}^{(n)} & {\bf 0}^{(n)} \\ \hline     
					 {\bf 0}^{(n)} & P_1^{(n)} \\ 
						\end{array}\right]\cdot \left[ \begin{array}{c|c}
					U_{11}^{(n)} & U_{12}^{(n)} \\ \hline     
					 U_{21}^{(n)} & U_{22}^{(n)} \\ 
\end{array}\right]=\\
&=\left[ \begin{array}{c|c}
					{\bf 0}^{(n)} & U_{21}^{(n)\dagger} P_1^{(n)} \\ \hline     
					 {\bf 0}^{(n)} & U_{22}^{(n)\dagger} P_1^{(n)} \\ 
						\end{array}\right]\cdot \left[ \begin{array}{c|c}
					U_{11}^{(n)} & U_{12}^{(n)} \\ \hline     
					 U_{21}^{(n)} & U_{22}^{(n)} \\ 
\end{array}\right]= \left[ \begin{array}{c|c}
U_{21}^{(n){\dagger}}P_1^{(n)}U_{21}^{(n)} & U_{21}^{(n){\dagger}}P_1^{(n)}U_{22}^{(n)} \\ \hline      U_{22}^{(n){\dagger}}P_1^{(n)}U_{21}^{(n)} & U_{22}^{(n){\dagger}}P_1^{(n)}U_{22}^{(n)} \\ 
\end{array}\right]=\\
&= \left[ \begin{array}{c|c}
I^{(n-1)}\otimes(U_{21}^{{\dagger}}P_1U_{21}) & I^{(n-1)}\otimes(U_{21}^{{\dagger}}P_1U_{22}) \\ \hline      I^{(n-1)}\otimes(U_{22}^{{\dagger}}P_1U_{21}^{(n)}) & I^{(n-1)}\otimes(U_{22}^{{\dagger}}P_1U_{22}) \\ 
\end{array}\right]=\Lambda_{U}^{(n+1)}
\end{eqnarray*}
Hence, $$\texttt P[^{\mathcal D}U_{[k;m,m+n]},\rho]=Tr[(I^{(m-1)}\otimes\Lambda_U^{(n+1)}\otimes I^{(k-m-n)})\rho]=Tr[(\Lambda_U^{(m+n)}\otimes I^{(k-m-n)})\rho].$$

\end{proof}
\end{theorem}
In this theorem we have considered the case where the binary operator $U$ is not a control-target operator. Accordingly with Def. \ref{Prob}, this assumption is essential in order to establish the correct expression of the probability of the circuit. In the case where the binary operator $U$ is a control-target gate, this expression will change, as showed in the following theorem. 
\begin{theorem}\ \\

Let $C{\tilde U}$ be a binary control-target operator $C{\tilde U}=\left[ \begin{array}{c|c}
					I & 0 \\ \hline     
					 0 & \tilde{U} \\ 
						\end{array}\right]$ (as represented in (\ref{CU})), with $\tilde U$ arbitrary unary gate. Let $\rho\in\mathcal D(\otimes^k\mathbb C^2).$ Then:
$$\texttt P[^{\mathcal D}C{\tilde U}_{[k;m,m+n]},\rho]=Tr[(\Lambda_{C{\tilde U}}^{(m+n)}\otimes I^{(k-m-n)})\rho],$$ where $\Lambda_{C{\tilde U}}^{(n+1)}=\left[ \begin{array}{c|c}
P_1^{(n)} & {\bf 0}^{(n)} \\ \hline      {\bf 0}^{(n)} & I^{(n-1)}\otimes(\tilde{U}^{\dagger}P_1\tilde{U}) \\ 
\end{array}\right].$

\begin{proof}\ \\

By Def.(\ref{Prob}), we have:
\begin{eqnarray*}
& & \texttt P[^{\mathcal D}C\tilde{U}_{[k;m,m+n]},\rho]=\\
&=& Tr[(I^{(m)}\otimes P_1^{(n)}\otimes I^{(k-m-n)})C\tilde{U}_{[k;m,m+n]}\,\,\rho\,\, (C\tilde{U}_{[k;m,m+n]})^{\dagger}] =\\
&=& Tr[((C\tilde{U}_{[k;m,m+n]})^{\dagger}(I^{(m)}\otimes P_1^{(n)}\otimes I^{(k-m-n)})C\tilde{U}_{[k;m,m+n]})\,\rho]=\\
 & & \text{(by Theorem (\ref{binary}))}
= Tr[(I^{(m-1)}\otimes  \left[ \begin{array}{c|c}
					I^{(n)} & {\bf 0}^{(n)} \\ \hline     
					 {\bf 0}^{(n)} & I^{(n-1)}\otimes \tilde{U}^{\dagger} \\ 
						\end{array}\right] \otimes I^{(k-m-n)})\cdot \\ 
&\cdot& (I^{(m)}\otimes  P_1^{(n)}  \otimes I^{(k-m-n)}) \cdot (I^{(m-1)}\otimes \left[ \begin{array}{c|c}
					I^{(n)} & {\bf 0}^{(n)} \\ \hline     
					 {\bf 0}^{(n)} & I^{(n-1)}\otimes \tilde{U} \\ 
						\end{array}\right]  \otimes I^{(k-m-n)}) \rho]=\\
&=& Tr[(I^{(m-1)}\otimes(\left[ \begin{array}{c|c}
					I^{(n)} & {\bf 0}^{(n)} \\ \hline     
					 {\bf 0}^{(n)} & I^{(n-1)}\otimes \tilde{U}^{\dagger} \\ 
						\end{array}\right]\cdot \left[ \begin{array}{c|c}
					P_1^{(n)} & 0 \\ \hline     
					 0 & P_1^{(n)} \\ 
						\end{array}\right]\cdot\\
& &\left[ \begin{array}{c|c}
					I^{(n)} & {\bf 0}^{(n)} \\ \hline     
					 {\bf 0}^{(n)} & I^{(n-1)}\otimes \tilde{U}\\ 
						\end{array}\right])\otimes I^{(k-m-n)})\,\rho].
\end{eqnarray*}

Let us notice that (by the \emph{mixed-produt property} of the tensor product):
\begin{eqnarray*}
&  \left[ \begin{array}{c|c}
					I^{(n)} & {\bf 0}^{(n)} \\ \hline     
					 {\bf 0}^{(n)} & I^{(n-1)}\otimes \tilde{U}^{\dagger} \\ 
						\end{array}\right]\cdot \left[ \begin{array}{c|c}
					P_1^{(n)} & 0 \\ \hline     
					 0 & P_1^{(n)} \\ 
						\end{array}\right]\cdot \left[ \begin{array}{c|c}
					I^{(n)} & {\bf 0}^{(n)} \\ \hline     
					 {\bf 0}^{(n)} & I^{(n-1)}\otimes \tilde{U}\\ 
						\end{array}\right]=\\
&=\left[ \begin{array}{c|c}
					P_1^{(n)} & {\bf 0}^{(n)}\\ \hline     
					 {\bf 0}^{(n)} & (I^{(n-1)}\otimes {\tilde U}^{\dagger}) P_1^{(n)} \\ 
						\end{array}\right]\cdot \left[ \begin{array}{c|c}
					I^{(n)} & {\bf 0}^{(n)} \\ \hline     
					{\bf 0}^{(n)} & I^{(n-1)}\otimes {\tilde U} \\ 
\end{array}\right]=\\
& \left[ \begin{array}{c|c}
P_1^{(n)} &  {\bf 0}^{(n)} \\ \hline       {\bf 0}^{(n)} & (I^{(n-1)}\otimes \tilde{U}^{\dagger})P_1^{(n)}(I^{(n-1)}\otimes \tilde{U}) \\ 
\end{array}\right]= \left[ \begin{array}{c|c}
P_1^{(n)} & {\bf 0}^{(n)} \\ \hline      {\bf 0}^{(n)} & I^{(n-1)}\otimes(\tilde{U}^{\dagger}P_1\tilde{U}) \\ 
\end{array}\right]=\Lambda_{C\tilde{U}}^{(n+1)}
\end{eqnarray*}

Hence, our claim.
\end{proof}
\end{theorem}

\begin{corollary}

$$\texttt P(\sqrt{Swap}_{[k;m,m+n]},\rho)=Tr[(P_1^{(m)}\otimes P_1^{(n)}\otimes I^{k-m-n})\rho].$$

\begin{proof}\ \\

By a straightforward calculation, it can be seen that, for an arbitrary $\rho\in\mathcal D(\otimes^k\mathbb C^2)$,  is: $$\Lambda_{\sqrt{Swap}_{[k;m,m+n]}}^{(n+1)}=\left[ \begin{array}{c|c} {\bf 0}^{(n)} & {\bf 0}^{(n)} \\ \hline     
					 {\bf 0}^{(n)} & P_1^{(n)} \\ 
\end{array}\right]=P_1\otimes P_1^{(n)}.$$

Hence, $\Lambda_{\sqrt{Swap}_{[k;m,m+n]}}^{(m+n)}=I^{(m-1)}\otimes \Lambda_{\sqrt{Swap}_{[k;m,m+n]}}^{(n+1)}=P_1^{(m)}\otimes P_1^{(n)}.$

By a direct application of Theorem (\ref{binary2}), easily follows our claim.
\end{proof}
\end{corollary}

Following a similar reasoning, it is easy to verify that $$\Lambda_{Swap_{[k;m,m+n]}}^{(n+1)}=\Lambda_{\sqrt{Swap}_{[k;m,m+n]}}^{(n+1)},$$ hence $\texttt P(Swap_{[k;m,m+n]},\rho)=\texttt P(\sqrt{Swap}_{[k;m,m+n]},\rho)$ for any $\rho\in\otimes^k\mathcal D(\mathbb C^2)$.

It is remarkable to notice that in the evaluation of these probability values, the gates $Swap$ and $\sqrt{Swap}_{[k;m,m+n]}$ plays the same role of the binary identity gate.

It is also easy to verify that the probability value of the $CNot_{[k;m,m+n]}$ gate in the context of the MT-QCL is in accord with the standard QCL.

As a final example concerning a ternary gate, we also calculate the probability of the generalized Fredkin gate.
\begin{theorem}\ \\

Let $\rho\in\mathcal D(\otimes^k\mathbb C^2)$ and let $F_{[k;m,m+n,m+n+l]}$ the $k$-dimensional Fredkin gate. We have that $$\texttt P(F_{[k;m,m+n,m+n+l]},\rho)=Tr[(I^{(m-1)}\otimes\Lambda_F^{(n+l+1)}\otimes I^{(k-m-n-l)})\rho],$$ where $\Lambda_F^{(n+l+1)}= P_0\otimes P_1^{(n)}\otimes P_1^{(l)}+P_1\otimes P_1^{(l)}\otimes P_1^{(n)}.$

\begin{proof}\ \\

The proof follows by considering the Fredkin gate $F_{[k;m,m+n,m+n+l]}$ (let's simply say $F$) as a Control-Swap gate, where the $m^{th}$ qubit plays the role of control and the $(m+n)^{th}$ and the $(m+n+l)^{th}$ qubits play the role of target. For this reason, in accord with Def.\ref{Prob}, the probability will be given by:

$$\texttt P(F,\rho)= Tr[(I^{(m)}\otimes P_1^{(n)})\otimes P_1^{(l)}\otimes I^{(k-m-n-l)}F\rho F].$$
Further, by considering the representation of the Fredkin gate as \cite{SF}: $$F^{(m,n,l)} = I^{m-1}(P_0 \otimes I^{(n+l)} + P_1\otimes Swap_{[n+l;n,n+l]}$$ and by a direct calculation, we obtain that $\texttt P(F_{[k;m,m+n,m+n+l]},\rho)=Tr[(I^{(m-1)}\otimes\Lambda_F^{(n+l+1)}\otimes I^{(k-m-n-l)})\rho],$ where 

\begin{eqnarray*}
& \Lambda_F^{(n++l+1)}=\\
& [P_0\otimes I^{(n+l)}+P_1\otimes Swap_{[n+l;n,n+l]}]\cdot[I\otimes(P_1^{(n)})\otimes P_1^{(l)}]\cdot\\
& [P_0\otimes I^{(n+l)}+P_1\otimes Swap_{[n+l;n,n+l]}]=\\
& = [P_0\otimes P_1^{(n)}\otimes P_1^{(l)}+P_1\otimes (Swap_{[n+l;n,n+l]}\cdot(P_1^{(n)}\otimes P_1^{(l)}))]\cdot\\
& [P_0\otimes I^{(n+l)}+P_1\otimes Swap_{[n+l;n,n+l]}]=\\
& (P_0\cdot P_0)\otimes((P_1^{(n)}\otimes P_1^{(l)})\cdot I^{(n+l)})+\\
& (P_0\cdot P_1)\otimes ((P_1^{(n)}\otimes P_1^{(l)})\cdot Swap_{[n+l;n,n+l]})+\\
& (P_1\cdot P_0)\otimes (Swap_{[n+l;n,n+l]}\cdot(P_1^{(n)}\otimes P_1^{(l)})\cdot I^{(n+l)}) +\\
& (P_1\cdot P_1)\otimes (Swap_{[n+l;n,n+l]}\cdot(P_1^{(n)}\otimes P_1^{(l)})\cdot Swap_{[n+l;n,n+l]}). 
\end{eqnarray*}
Hence our claim.
\end{proof}
\end{theorem}

The results provided in this section permit to realize that the QCL can be seens as a particular (one-target) generalization of the MT-QCL. Indeed, if we confine to the set of the one-target qubits, the MT-QCL replace the same probabilistic results of the standard QCL; on the other hand, the MT-QCL allows to consider also multi-target gates that are not possible to take into account in the context of the standard QCL.

\section{Conclusion and further developments}
In this paper we have introduced the formal framework of a generalization of the standard Quantum Computational Logic, that takes into account the different roles of the qubits in a quantum circuit. The most relevant utility of the Quantum Computational Logic is based on the fact that it turns out to be strongly holistic (non compositional) and it is estremely useful to represent all these kinds of situations where the meaning of a compound system (logically represented by a sentence) is not simply dependent on the meaning of its subsystems (logically represented by the atomic senteces), but it has to be considere as a whole. Further, the standard Quantum Computational Logic has also the extreme privilege to be \emph{context dependent}, i.e. the same sentence can assume different meanings in different contexts. Non-compositionality and context-dependences are two features that make the Quantum Computational Logic extremely useful to describe situations related to very different aspects of the real rational activity in many different contexts such as common language, human psychology, machine learning and even the usual way to perceive the music \cite{BDCGLS,DGLNS,DGLS,HSFP,SSDMG}. Let us notice how the QCL can reach a so strong expressive power by considering in the language only a particular class of gates (one-target gates). The Multi Target Quantum Computational Logic hereby introduced, allows to consider infinite many more gates (i.e. logical connectives) inspired by the standard language of Quantum Computational Logic and, at the same time, it is more realistic in accord with what happens in a real quantum computational process. On this basis, this expansion of the language can be considered as a promising tool to boost the expressive power of the QCL. As a natural pursuance, the investigation on the theoretical and semantic benefits of this more expressive representation seems to be an interesting argument for further developments. In particular, future investigations will be devoted - from a theoretical point of view - to a complete study on the algebraic properties of this new logical structures and - from a more applicative viewpoint - to continue to investigate on the benefit obtained by applying these logical structures inspired by quantum theory to non-standard contexts such as human behavior, machine learning and so on.

\begin{acknowledgements}
This work is partially supported by Regione Autonoma della Sardegna within the project ``\emph{Time-logical evolution of correlated microscopic systems}" CRP 55, L.R. 7/2007 (2015).
\end{acknowledgements}



\end{document}